\documentstyle[aps,preprint,floats,tighten]{revtex}

\input epsf
\def\plotone#1{\centering \leavevmode
\epsfxsize= 1.0\columnwidth \epsfbox{#1}}

\def\hatn{{\bf \hat n}}

\newcommand{\be}{\begin{equation}}
\newcommand{\ee}{\end{equation}}
\newcommand{\bea}{\begin{eqnarray}}
\newcommand{\eea}{\end{eqnarray}}
\def\pslash{{\cal P}{\hbox{\kern-6pt $\slash$}}}

\long\def\comment#1{}

\input{psfig.tex}
\begin{document}
\draft
\title{Aspects of the Cosmic Microwave Background Dipole\footnote{Dedicated to
     the memory of David Wilkinson.}}
\author{Marc Kamionkowski$^1$ and Lloyd
Knox$^2$}
\address{$^1$Mail Code 130-33, California Institute of Technology,
Pasadena, CA 91125}
\address{$^2$Department of Physics, University of California, Davis,
CA 95616}
\date{September 2002}
\maketitle

\begin{abstract}
Cosmic microwave background (CMB) experiments generally infer a
temperature fluctuation from a measured intensity fluctuation
through the first term in the Taylor expansion of the Planck
function, the relation between the intensity in a given frequency
and the temperature.  However, with the forthcoming Planck
satellite, and perhaps even with the Microwave Anisotropy Probe,
the CMB-dipole amplitude will be large enough to
warrant inclusion of the next higher order term.  To quadratic
order in the dipole amplitude, there is an intensity quadrupole
induced by the dipole
with a frequency dependence given by the second derivative of
the Planck function.  The Planck satellite should be able to
detect this dipole-induced intensity quadrupole and distinguish
it through its frequency depdendence from the intrinsic CMB
temperature and foreground quadrupoles.  This higher-order
effect provides a robust pre-determined target that may provide tests
of Planck's and MAP's large-angle-fluctuation measurements and
of their techniques for multi-frequency foreground subtraction.
\end{abstract}

\pacs{PACS number(s):}

\section{Introduction}
\label{sec:intro}

The primary aim of NASA's recently launched Microwave Anisotropy
Probe (MAP) \cite{MAP} and the European Space Agency's Planck
satellite \cite{Planck} will be to measure small-scale
fluctuations in the cosmic microwave background (CMB) in order
to better determine cosmological parameters \cite{Knox,jkks} and
test inflationary cosmology \cite{KamKos}, and thus 
improve upon the already remarkable results from
recent balloon and ground experiments \cite{toco}.
However, the most salient feature in the CMB is not this small-scale
structure; it is the much more prominent dipole
\cite{dipole} of amplitude $\Delta T=3.365\pm0.027$ mK.

The simplest explanation for the dipole is a Local Group
velocity $v=627\pm22$ km~sec$^{-1}$ toward
$(l,b)=(276^\circ\pm3^\circ,33^\circ\pm3^\circ)$ with respect to
the CMB rest frame.  Some measurements of galaxy
velocities on large scales suggest independently, although far
less precisely, a similar value \cite{riess}.  It has thus been
generally accepted that the CMB dipole is due to our peculiar
velocity, but theorists have occasionally speculated that the
dipole might be due, at least in part, to an intrinsic
temperature fluctuation \cite{intrinsic}.

If we had sufficiently precise multi-frequency observations to
measure the frequency spectrum at each point on the sky, it
would alway be consistent with a blackbody spectrum, although
with an angle-dependent temperature.  In practice, the
observation frequencies and sensitivities are limited.  The
temperature at each point on the sky is thus determined
by measuring the intensity fluctuation, and then converting it
to a temperature fluctuation, generally assuming
that the temperature fluctuation is sufficiently small that the
intensity fluctuation can be related to the temperature
fluctuation by the first term in a Taylor expansion of the
Planck function, the relation between the temperature and
intensity.  With prior experiments, the temperature fluctuation
has always been sufficiently small that this approximation has
been warranted.

In this paper we point out that with the observed dipole
amplitude and Planck's (and possibly MAP's) sensitivity,
this will no longer be a good approximation---the second-order
term in the Taylor expansion will be non-neglible.   To
quadratic order in the dipole amplitude, there is an {\it
intensity} quadrupole with a frequency dependence given by the
second derivative of a Planck function.  Since this frequency
dependence differs from that of the first-order term, this
dipole-induced intensity quadrupole can be disentangled from the
temperature quadrupole, even after taking into account a number
of foregrounds, as we show below.  This higher-order effect will
provide a robust pre-determined signal that Planck and possibly
MAP should be able to detect.  It can be used to calibrate
large-angle-fluctuation measurements and.or benchmark
multi-frequency foreground-subtraction techniques.  It may thus
be a useful addition to the astrophysical point sources and
annual modulation of the dipole that have until now served as
calibration sources for CMB experiments.

In the next Section we discuss the dipole-induced intensity
quadrupole, which we refer to simply as the ``dipole
quadrupole'' and abbreviate DQ.  In Section III, we then show
that the DQ should be detectable by Planck and distinguishable
from the CMB quadrupole, even after subtracting several
foregrounds.  Our analysis in Sections II and III assumes that
the dipole is due entirely to the velocity of the solar system
with respect to the CMB rest frame.  To indicate the magnitude
and detectability of of the DQ, we evaluate how well the
magnitude and orientation of the peculiar velocity could be
determined---assuming the dipole were due to a peculiar
velocity---from the DQ alone.  In Section IV we
show that the same DQ arises even if the CMB dipole is an
intrinsic temperature fluctuation, rather than the result of a
peculiar velocity.  Section V provides some closing remarks.

\section{The Dipole-Induced Intensity Quadrupole}
\label{sec:formalism}

The specific intensity of a blackbody of temperature $T_{\rm
CMB}$ is
\begin{equation}
     I'_{\nu'}= C{ x'^3 \over e^{x'}-1},
\end{equation}
where $C=2 (k_B T_{\rm CMB})^3 (hc)^{-2}$ is a constant, 
$x'=h \nu'(k_B T_{\rm CMB})^{-1}$, $\nu'$ is the photon frequency, and
$h$, $k_B$, and $c$ are, respectively, the Planck and Boltzmann
constants and the speed of light.  The photon frequency
measured by an observer moving with velocity $v$ relative
to the blackbody is $\nu=\gamma\nu'(1+\beta\mu)$, where
$\beta=v/c$, $\gamma=(1-\beta^2)^{-1/2}$, and $\mu$ is the
cosine of the angle between the velocity and
the photon direction.  The specific
intensity transforms as $I'_{\nu'}/\nu'^3=I_\nu/\nu^3$, and thus
the observer-frame specific intensity is
\begin{equation}
     I_\nu = C {x^3 \over e^{x \gamma (1+\beta\mu)}-1},
\end{equation}
where $x=h\nu (k_B T_{\rm CMB})^{-1}$.  Expanding in $\beta$,
\be
I_\nu = C {x^3 \over e^x-1} \left\{ 
\left[ 1+ O(\beta^2)\right] -
     f(x) \beta \mu  
     +  f(x) g(x) \beta^2(\mu^2-1/3) + \cdots \right\}.
\label{eqn:Iquad}
\ee
where $f(x)\equiv x e^x(e^x-1)$ and $g(x)\equiv(x/2)\coth(x/2)$.

The term linear in $\beta$ is the dipole, with the appropriate
frequency spectrum $f(x)$ for the dipole \cite{dipolefour}, which is
also more generally the frequency spectrum for a thermal
fluctuation.  However, to order $\beta^2$, there is another
term, the DQ, with a frequency spectrum that differs from that
from small thermal fluctuations.  With COBE-{\sl DMR}'s
sensitivity and frequencies (in the Rayleigh-Jeans
regime), the frequency dependence was negligible.  However, with
the improved
sensitivity and expanded frequency range of new satellite
experiments, the frequency dependence will become detectable.
This frequency dependence was noted in
Refs. \cite{italians}, but seems to have since been
overlooked.  Refs. \cite{audit} identified
the effect (sometimes referred to as the ``kinematic quadrupole'') in
calculations of CMB polarization induced by reionization, but
they did not consider the CMB dipole.  Ref. \cite{challinortwo}
discusses peculiar-velocity effects on the CMB, but only on
small-scale fluctuations.

\begin{figure}[htbp]
    \plotone{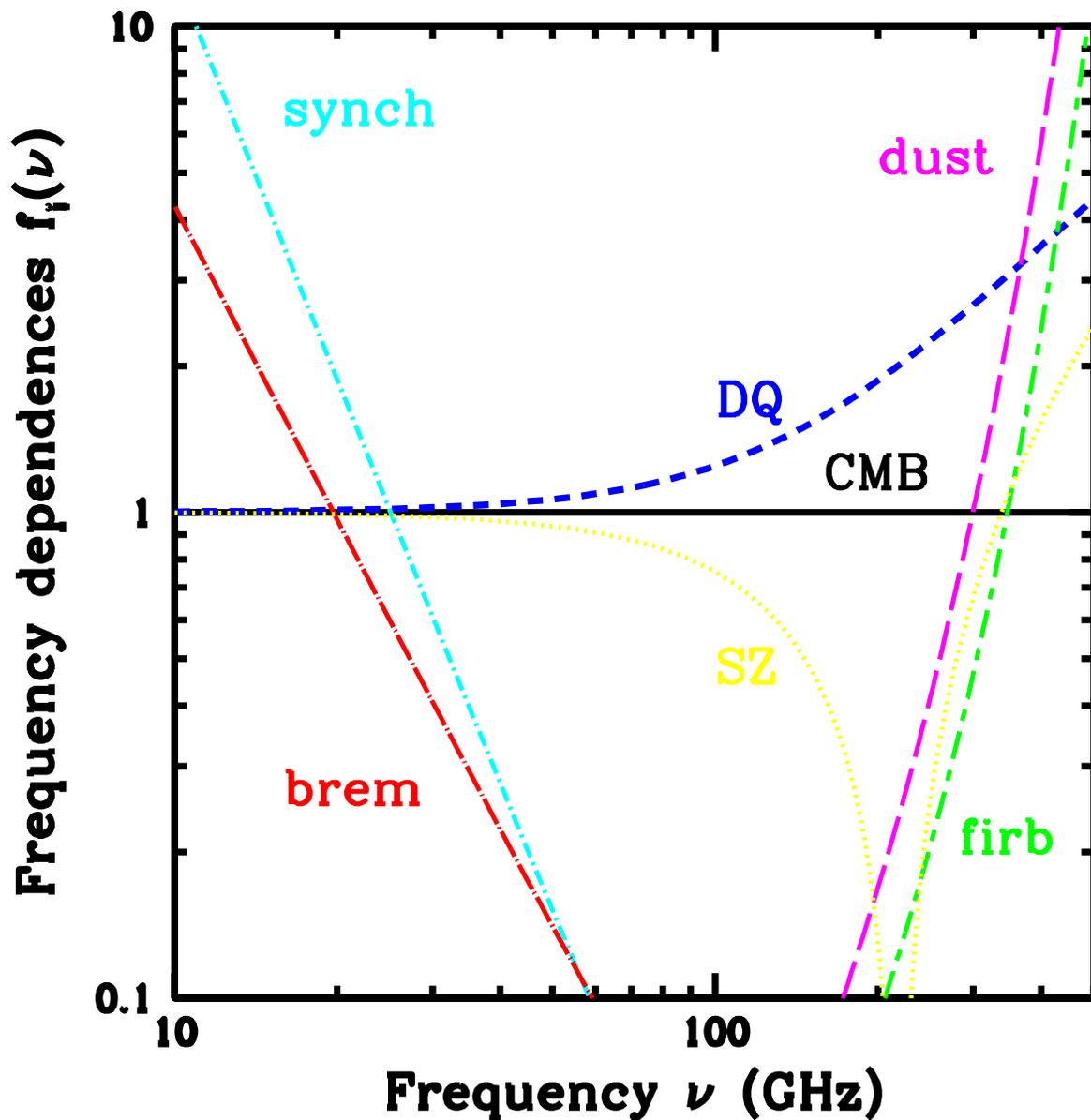}
\caption{The frequency dependence of the cosmological (solid;
     black), kinematic (short-dash; blue), dust (long-dash;
     magenta), synchrotron (dot--short-dash; cyan),
     bremsstrahlung (dot--long-dash; red), FIRB
     (short-dash--long-dash; green), and Sunyaev-Zeldovich
     quadrupole moments, all referenced to the frequency
     dependence of the cosmological quadrupole.  For the
     Sunyaev-Zeldovich effect, we plot the absolute value as the
     frequency dependence changes sign near $\nu=220$ GHz.  The
     amplitudes are arbitrary, as our analysis assumes that they
     will be determined by the data.}
\label{fig:nuplot}
\end{figure}

\section{Detectability of the DQ}
\label{sec:detecability}

To assess the detectability of the DQ, we determine how well the
amplitude and direction of the solar-system velocity can be
determined with MAP and Planck under the assumption that the
velocity is to be reconstructed entirely from the DQ.

Since the frequency dependences of the CMB quadrupole and the DQ
differ, they can be distinguished if the quadrupole moments of
the intensity are determined in several frequencies.  However,
the CMB quadrupole and the DQ will also have to be distinguished
from the quadrupole moments due to unsubtracted, or imperfectly
substracted, foregrounds, each with its own frequency dependence.

In what follows we discuss the brightness temperature (which is
proportional to the intensity) rather than the intensity so that
we work with units that are familiar in CMB studies.
If the brightness temperature
$T_\nu(\hatn)$ is measured in several frequencies $\nu$ as a
function of position $\hatn$ over the entire sky, then the five (for
$m=-2,-1,0,1,2$) quadrupole components $T_{m,\nu}$
can be constructed from $T_{m,\nu}=\int Y_{2m}(\hatn)
T_\nu(\hatn) d\hatn$, where $Y_{lm}(\hatn)$ are spherical
harmonics.  The possible contributions to
these $T_{m,\nu}$ from several unsubtracted foregrounds (dust,
synchrotron radiation, bremsstrahlung, the far infrared
background, and the Sunyaev-Zeldovich effect), which will have
frequency dependences shown in Fig.~\ref{fig:nuplot}, must also
be considered.  The
quadrupole moment at each frequency is then the sum
of the contributions from all of these sources:
$T_{m,\nu}=\sum_i T_{m}^{i} f_i(\nu)$, where
$i=\{{\rm CMB,DQ,dust,synch,brem,firb,SZ}\}$, and $f_i(\nu)$ is the
frequency dependence of source $i$, referenced to the CMB
frequency dependence.  Thus, $f_{\rm CMB}(\nu)=1$; $f_{\rm
DQ}(\nu)=(x/2) \coth(x/2)$; $f_{\rm dust}(\nu)=x^2
e^{x_d-x}(e^x-1)^2 (e^{x_d}-1)^{-2}$ (where $x_d=h \nu (k_B
T_{\rm dust})^{-1}$ and we have taken a thermal dust spectrum with
$T_{\rm dust}=20$ K and emissivity index $\alpha=2$); $f_{\rm
synch}(\nu)=x^{-4.8}(e^x-1)^2 e^{-x}$;
$f_{\rm brem}(\nu)=x^{-4.16}(e^x-1)^2 e^{-x}$;  $f_{\rm firb}(\nu)=x^{1.64}
e^{x_f-x}(e^x-1)^2 (e^{x_f}-1)^{-2}$ (where $x_f=h \nu (k_B
T_{\rm firb})^{-1}$ and we have taken a thermal FIRB spectrum with
$T_{\rm firb}=18.5$ K and emissivity index $\alpha=0.64$ as a fit
to data \cite{fixsen98}); and
$f_{\rm SZ}(\nu)=2 f_{\rm CMB}(\nu)-f_{\rm DQ}(\nu)$. For a review
of foreground properties see Ref.~\cite{tegmark00}.

For each of the five components $m$, we will have five (for the MAP
frequencies of 22, 30, 40, 60, and 90 GHz) or nine (for the
Planck frequencies of 30, 44, 65, 100, 143, 217, 353, 545, and
857 GHz) data points $T_{m,\nu}$ that we
model with a vector of parameters $s_i=T_{m}^i$ again for
$i=\{{\rm CMB,DQ,dust,synch,brem,firb,SZ}\}$. 
Following Ref. \cite{jkks}, the standard error ($1\sigma$) to
the quadrupole component $T_{m}^{\rm DQ}$ will be $\sigma_{m}^{\rm
DQ}= [\alpha^{-1/2}]_{\rm DQ,DQ}$, where 
\begin{equation}
     \alpha_{ij} = \sum_\nu \left[ {\partial T_{m,\nu} \over
     \partial s_i} {\partial T_{m,\nu} \over
     \partial s_j}\right] {1\over \sigma_{m,\nu}^2} = \sum_\nu
     {f_i(\nu) f_j(\nu) \over \sigma_{m,\nu}^2},
\end{equation}
is the covariance (Fisher) matrix.  Here, $\sigma_{m,\nu}^2$ is
the variance to the quadrupole from instrumental noise at
frequency $\nu$; we use 4-year values for MAP from
Ref. \cite{MAP} and 2-year values for Planck from Ref. \cite{Planck}.

Once the DQ components $T_m^{\rm DQ}$ have
been determined, the peculiar-velocity components can be found.
Expanding the angular dependence, $(\mu^2-1/3)$, where $\mu$ is
the cosine of the angle between the velocity and the photon
direction, in spherical harmonics gives us $(T_{m=0}^{\rm
DQ}/T_{\rm CMB})=-\sqrt{4 \pi/45}(\beta_x^2+\beta_y^2-2\beta_z^2)$,
$\sqrt{2} ({\rm Re} T_{m=1}^{\rm DQ}/T_{\rm CMB})=-\sqrt{16\pi/15} \beta_x
\beta_z$, $\sqrt{2} ({\rm Im} T_{m=1}^{\rm DQ}/T_{\rm CMB})=-\sqrt{16\pi/15}
\beta_y \beta_z$,  $\sqrt{2} ({\rm Re} T_{m=2}^{\rm
DQ}/T_{\rm CMB})=-\sqrt{4\pi/15} (\beta_x^2 - \beta_y^2)$, and  $\sqrt{2}
({\rm Im} T_{m=2}^{\rm DQ}/T_{\rm CMB})=-\sqrt{16\pi/15}
\beta_x \beta_y$.  Let us suppose that these moments have been
measured precisely.  Then the equations for $T_{m=0}^{\rm
DQ} (\equiv a)$, ${\rm Re} T_{m=2}^{\rm DQ}(\equiv b)$, and
${\rm Im} T_{m=2}^{\rm DQ}(\equiv c)$ can be inverted (after
introducing the shorthand $a$, $b$, and $c$) to give us the
components $\beta_x^2=\sqrt{15(16\pi)^{-1}}(\sqrt{b^2+c^2}+b)$,
$\beta_y^2=\sqrt{15(16\pi)^{-1}}(\sqrt{b^2+c^2}-b)$, and 
$\beta_z^2=\sqrt{15(16\pi)^{-1}} (\sqrt{b^2+c^2}-\sqrt{3} a)$.  Since
we have now determined the squares of three components of the
velocity, we are still left with a residual eightfold degeneracy
in the velocity.  However, the signs of the components ${\rm Re}
T_{m=1}^{\rm DQ}$ and ${\rm Im} T_{m=1}^{\rm DQ}$, which are
proportional, respectively, to $\beta_x\beta_z$ and $\beta_y\beta_z$, 
can be used to
break two of these degeneracies, leaving us with only a twofold
degeneracy, the sign of the velocity, which is undetermined by
the quadrupole.

When the measurements are done, we will then want to check
whether the peculiar velocity induced from the DQ is consistent
with that inferred from the dipole itself.  If the data analysis
and our understanding of the instrument are reliable, then the
two velocities should be consistent.  To quantify the degree to
which MAP and Planck can test this consistency we calculate,
assuming $v=0$, the expectation value of $\chi^2$ for the 
hypothesis that $v = 627$ km~sec$^{-1}$.  If, e.g.,
$\chi^2=9$ then the $v=0$ hypothesis can be distinguished from the
$v=627$ km~sec$^{-1}$ hypothesis at $3\sigma$.  We find, for
arbitrary $v$, $\chi^2 = 9 (4.2~\mu{\rm K}/\sigma_m^{\rm DQ})^2
\left[v/ (627{\rm ~km~sec}^{-1}) \right]^4$.  From this $\chi^2$
we also derive the minimum $v$ distinguishable at $3\sigma$ from $v=0$
to be $v_{\rm min} = 305 $ km~sec$^{-1}$   $(\sigma_m^{\rm
DQ}/\mu{\rm K})^{1/2}$.

As another indicator of the precision with which the DQ can be
reconstructed, we now determine the precision with which the
magnitude and orientation of the velocity can be inferred from
the DQ alone.
There will be a measurement error of $\sigma_m^{\rm
DQ}$ to each of the $T_m^{\rm DQ}$, as described above, and
the three velocity components, $v_x$, $v_y$, and $v_z$, will be
fit to all five of these quadrupole components.  
To estimate the errors on these components
we once again calculate a Fisher matrix, this time for these
five data points and the parameters $v_x$, $v_y$, and $v_z$.
We then invert the Fisher matrix and choose the $z$
axis along the inferred velocity so that $v =v_z$.  We then find that the
measurement error (at $3\sigma$) to the 
velocity is $\Delta v=3 \sqrt{45(64\pi)^{-1}} \sigma_m^{\rm DQ}
c^2 ( v_z T_{\rm CMB})^{-1}$, where $v$ is the best-fit velocity.
Under the null hypothesis that the peculiar velocity is that
inferred from the dipole, we find $\Delta v=225(\sigma_m^{\rm
DQ}/\mu{\rm K})$ km~sec$^{-1}$.  The Fisher-matrix analysis
tells us that the measurement errors to the $x$ and $y$
components (those perpendicular to the best-fit velocity) are
each $\sqrt{4/3}$ times that for the $z$ component.  Thus, under
the same null hypothesis, the ($3\sigma$) error to the velocity
orientation will be $24^\circ(\sigma_m^{\rm DQ}/\mu{\rm K})$.

\begin{table*}[htbp]
\caption{Values for detectability of the dipole-induced
     intensity quadrupole.  The $\sigma_m^{\rm DQ}$ column gives the standard
     error to the DQ amplitude.  The $\chi^2$ column quantifies the
     ability to distinguish 
     $v=0$ from $v= 627$ km~sec$^{-1}$ using the DQ.  
     The quadrupole components that are assumed to be fit to the multifrequency
     data are the DQ, CMB quadrupole (T),
     dust (D), synchrotron radiation (Synch), bremsstrahlung
     (Brem), far infrared background (FIRB), and Sunyaev-Zeldovich
     effect (SZ).  The ``NA'' in the $\chi^2$ column
     indicates that if the SZ effect is taken into account, our
     covariance matrix formally gives an infinite result
     because of the degeneracy between the frequency dependences
     of the CMB, the DQ, and Sunyaev-Zeldovich
     effect.  The column $\Delta v$ gives the error
     (at $3\sigma$) to the magnitude of the velocity. 
     We list NA for $\Delta v$ for the cases where the DQ
     is unlikely to be detected.  The last column gives the
     anticipated ($3\sigma$) error to the orientation of the
     velocity.}
\begin{center}
\begin{tabular}{cccccc}
experiment     &     components      &  $\sigma_m^{\rm DQ}$
($\mu$K) &  $\chi^2$  &
$\Delta v$ (km~sec$^{-1}$) & $\Delta \theta$ (degrees) \\ \hline \hline
MAP   &         DQ                        & 0.04 & $10^5$ & 9 & 0.9 \\  \hline
MAP    &        DQ$+$T                    & 0.6 & 440 & 140 & 14 \\  \hline
MAP    &        DQ$+$T$+$Synch$+$Brem     & 1.7 & 52 & 390 & 42  \\  \hline
MAP    &        DQ$+$T$+$D                & 16 & 0.66 & NA & NA \\  \hline
MAP    &        DQ$+$T$+$D$+$Synch$+$Brem & 73 & 0.032 &  NA & NA \\  \hline
Planck  &      DQ                & 0.005&  $6\times 10^6$ &  1.1 & 0.12 \\  \hline
Planck  &      DQ$+$T            & 0.02 & $3\times 10^5$   & 4.9 & 0.52 \\  \hline
Planck  &      DQ$+$T$+$D        & 0.02 & $3\times 10^5$ & 5.0  & 0.53 \\  \hline
Planck  &      DQ$+$T$+$D$+$FIRB   & 0.036 & $10^5$       & 8.1  &  0.86 \\  \hline
Planck	&      DQ$+$T$+$D$+$FIRB$+$SZ     & NA & A & NA  &  NA \\  \hline
Planck  &      DQ$+$T$+$D$+$FIRB$+$Synch$+$Brem & 0.2 & 4,000 & 45 & 4.8 \\  \hline
\end{tabular}
\end{center}
\label{tab:values}
\end{table*}

Table \ref{tab:values} shows results of our calculations of
$\chi^2$ and $\Delta v$
assuming a variety of combinations of components will be fit to
the data.  If the dust contribution to MAP
must be determined from the MAP data itself, then the smallest
detectable peculiar velocity is $\sim2100$ km~sec$^{-1}$,
too big to be interesting.  This is because
the highest-frequency channel becomes a dust
monitor, and the remaining MAP channels at $\lesssim60$ GHz do not
provide enough leverage to disentangle the DQ and the CMB
quadrupole.  However, if the dust quadrupole can be
determined precisely from other observations, then MAP might be
able to isolate the DQ to better than $7\sigma$.  The synchrotron and
bremsstrahlung foregrounds are strongest at low
frequencies, where the DQ and CMB frequency dependences
are similar.  Thus, if synchrotron and bremsstrahlung emission
are included in the analysis, the lowest-frequency channels
become foreground monitors and the ability to separate the
DQ is not degraded significantly.

Because of the improved detector sensitivity, and
especially the broadened frequency coverage, the outlook for
Planck is much better.  Even without the highest-frequency
channels, which act as dust monitors, there is still
a sufficiently broad spectrum of frequencies $\gtrsim100$ GHz
where the spectral dependences of the DQ and the CMB
quadrupole differ the most.  Again, the synchrotron and
bremsstrahlung foregrounds contribute mostly at low frequencies
and thus do not degrade significantly the DQ
signal.  Quite remarkably, even if we marginalize over a number
of uncertain foreground amplitudes, Planck should be able to
detect the DQ as long as the velocity is
greater than roughly 140 km~sec$^{-1}$.  Moreover, Planck should
be able to detect a deviation from the dipole-inferred velocity
as small as 45 km~sec$^{-1}$ and determine its direction (modulo
the sign) to better than $5^\circ$.

It turns out that the Sunyaev-Zeldovich frequency dependence
is a linear combination of the CMB-quadrupole and
DQ frequency dependences.  Thus, the DQ is formally degenerate
with the cosmological and Sunyaev-Zeldovich quadrupoles.
However, the quadrupole moment due to the Sunyaev-Zeldovich
effect will probably be negligible.  Calculations of the
Sunyaev-Zeldovich power spectrum near $l\simeq100$ find $l^2
C_l/(2\pi) \lesssim 10^{-12}$ \cite{springel}.  Doing shot-noise
extrapolation to the quadrupole, we find
$\sigma_{m}^{\rm SZ} \lesssim 0.15\, \mu{\rm K}$, corresponding
to a velocity uncertainty (at $3\sigma$) $\lesssim 35$
km~sec$^{-1}$, which is just a bit below the expected
statistical uncertainty indicated in Table \ref{tab:values} for
Planck.  The  actual SZ number will probably be much smaller, and
certainly much more will be known by the time Planck flies.  The SZ
octupole will be useful in constraining the possible SZ quadrupole
contribution.

Our foreground modeling makes several simplifying assumptions.
For example, we do not actually
know the frequency dependences perfectly and they can 
vary spatially (due to, e.g., spatial variation in chemical
composition of the dust).  There may also be components
we have not yet considered, such as spinning dust.  Further,
about 25\% of the sky will be lost to a Galactic-plane cut.
It is thus possible that Planck may not be able to achieve the
tiny velocity errors quoted in the table.  However, the 
values in the final row may be achievable.  Although
uncertain, the frequency dependences of the foregrounds are all
(with the exception of the SZ effect) considerably different than those
of the the DQ and the CMB quadrupole.  The foregrounds are
all likely to have low amplitudes near 100 GHz:
$\sim 3\,\mu$K for dust, $\sim 1\,\mu$K  for bremsstrahlung and
less for everything else \cite{kogut96}.  Thus the DQ
is not far below foreground quadrupoles, so multi-frequency
foreground subtraction need not be done to better than
about 10\%.  Of course, only the measurements themselves will
answer these questions definitively.

\section{An Intrinsic Dipole?}

Although a peculiar velocity is the simplest explanation for the
dipole, and certainly that most consistent with the prevailing
inflationary paradigm, it is also possible that the dipole
could be due, at least in part, to
an intrinsic temperature dipole produced by a super-horizon
entropy perturbation \cite{intrinsic}.  In this Section, we show
that even if the temperature dipole is intrinsic, the same DQ
still arises.  We also clarify the distinction between an
intrinsic CMB temperature fluctuation, the thermal quadrupole
induced by a peculiar velocity, and the DQ, which unlike the
other two, is {\it not} a temperature quadrupole.

A velocity $v$ induces a temperature pattern $T(\theta) =
T_{\rm CMB}(1+\beta \mu)^{-1}\sqrt{1-\beta^2}$ \cite{pw}.  To
second order in $\beta^2$, this can be written as
\be
     T(\theta)=T_{\rm CMB}\{[1+O(\beta^2)] - \beta\mu + \beta^2
     (\mu^2-1/3) \}.
\label{eqn:Tquad}
\ee
Thus, the peculiar velocity induces a temperature quadrupole of
magnitude $O(\beta^2 T_{\rm CMB})$, in addition to the dipole of
amplitude $O(\beta T_{\rm CMB})$.  However, the quadrupole
in the {\it intensity} [Eq. (\ref{eqn:Iquad})] arises from a
combination of both the {\it temperature} dipole and the
temperature quadrupole.  To see this, we rewrite
Eq. (\ref{eqn:Tquad}) as 
\be
     T(\theta)=T_{\rm CMB}\{[1+O(\beta^2)] - \beta_1\mu +
     \beta_2^2 (\mu^2-1/3) \},
\label{eqn:Tquadprime}
\ee
so that we can see where the DQ comes from.
Doing so, we find that Eq. (\ref{eqn:Iquad}) becomes
\be
     I_\nu = C {x^3 \over e^x-1} \left\{ 
     \left[ 1+ O(\beta^2)\right] -
          f(x) \beta_1 \mu  
	       +  f(x) \left[\beta_2^2 +\beta_1^2
	       \left(g(x)-1\right)\right] (\mu^2-1/3)
	       + \cdots \right\}. 
\label{eqn:Iquadprime}
\ee
We thus see that what we have been calling the DQ consists of
two parts:
The first is the $\beta_2^2$ term, which is due to the
{\it temperature} quadrupole induced by our peculiar velocity.  Since
this is an honest-to-goodness temperature quadrupole, it has the
frequency dependence of the usual lowest-order thermal
fluctuation, and it cannot be distinguished from an intrinsic 
CMB temperature fluctuation.  The second term, proportional to
$\beta_1^2$, arises from the term in the Taylor expansion of the
intensity that is second order in the dipole amplitude.  This
second-order term has a frequency dependence that differs from
the usual lowest-order thermal fluctuation.  

It is also clear from Eqs. (\ref{eqn:Tquadprime}) and
(\ref{eqn:Iquadprime}) that even if the temperature dipole were
intrinsic (that is, due to an entropy perturbation), then there
would still be a DQ.  Although the frequency dependence is
$g(x)-1$ rather than $g(x)$, only the part 
proportional to $g(x)$ can be distinguished by multi-frequency
maps from a temperature quadrupole.  For this reason, the
amplitude and orientation of the DQ would be exactly the same as
if the dipole were due to a velocity, and the detectability
would be exactly as determined above.  Likewise, the DQ can{\it
not} be used to tell whether the dipole is due to a peculiar
velocity or due to an entropy perturbation, as previously
suggested \cite{italians}.\footnote{We thank E. Wright for
illuminating discussions on this point and for pointing out an
error in an earlier draft.}

Finally, we mention one last point.  Strictly speaking, an
intrinsic CMB temperature quadrupole will produce a contribution
to the intensity quadrupole with a frequency dependence
proportional to $g(x)$, just like the DQ.  However, if the
temperature-quadrupole amplitude is $\Delta T$ ($\sim10^{-5}$
for the intrinsic fluctuation and $\sim10^{-6}$ for the
velocity-induced temperature quadrupole), then this contribution
will be of order $(\Delta T)^2 \lesssim10^{-10}$, much smaller
than the $10^{-6}$ expected for the DQ.

\section{Conclusions}

We have shown that the dipole amplitude is sufficiently large
that the discrepancy between the exact frequency dependence of a
thermal fluctuation and the lowest-order frequency dependence
usually assumed will be detectable by Planck and possibly MAP.
To second-order in the dipole amplitude, this discrepancy is
manifest as an intensity quadrupole that can be distinguished
with multi-frequency measurements from an intrinsic temperature
quadrupole.  This provides a robust pre-determined target for
CMB experiments and it may prove to be a useful tool for
calibration for forthcoming space experiments.

\smallskip
We thank the Santa Barbara KITP for hospitality.  This work was
supported at Caltech by NSF AST-0096023, NASA NAG5-9821, and DoE
DE-FG03-92-ER40701, at Davis by NASA NAG5-11098, and at the KITP
by NSF PHY99-07949.


\begin{thebibliography}{99}

\bibitem{MAP} The MAP Collaboration: {\tt http://map.gsfc.nasa.gov}.

\bibitem{Planck} The Planck Collaboration:
     {\tt http://sci.esa.int/planck}.

\bibitem{Knox} L. Knox, {\sl Phys. Rev. D} {\bf 52}, 4307
     (1995).

\bibitem{jkks} G. Jungman, M. Kamionkowski, A. Kosowsky, and
     D. N. Spergel, {\sl Phys. Rev. D} {\bf 54}, 1332
     (1996).

\bibitem{KamKos} For a review, see, e.g., M. Kamionkowski and A. Kosowsky,
     {\sl Annu. Rev. Nucl. Part. Sci.} {\bf 49}, 77 (1999).

\bibitem{toco} A. D. Miller et al., {\sl Astrophys. J. Lett.}
     {\bf 524}, L1 (1999); P. de Bernardis et al.,
     {\sl Nature} {\bf 404}, 955 (2000); S. Hanany et al.,
     {\sl Astrophys. J. Lett.} {\bf 545}, L5 (2000);
     N. W. Halverson et al., {\sl
     Astrophys. J.} {\bf 568}, 38 (2002); B. S. Mason et al.,
     astro-ph/0205384.

\bibitem{dipole} G. F. Smoot, M. V. Gorenstein, and
     R. A. Muller, {\sl Phys. Rev. Lett.} {\bf 39}, 898 (1977);
     E. S. Cheng, P. R. Saulson, D. T. Wilkinson, and
     B. E. Corey, {\sl Astrophys. J. Lett.} {\bf 232}, L139
     (1979); P. M. Lubin, G. L. Epstein, and G. F. Smoot,
     {\sl Phys. Rev. Lett.} {\bf 50}, 616 (1983); D. J. Fixsen,
     E. S. Cheng, and D. T. Wilkinson, {\sl Phys. Rev. Lett.}
     {\bf 50}, 620 (1983); A. Kogut et al., {\sl Astrophys. J.}
     {\bf 419} 1 (1994).

\bibitem{riess} A. G. Riess, W. H. Press, and R. P. Kirshner,
     {\sl Astrophys. J. Lett.} {\bf 445}, L91 (1995).

\bibitem{intrinsic} L. P. Grishchuk and Ya. B. Zeldovich,
     {\sl Sov. Astron.} {\bf 22}, 125 (1978); D. Langlois, {\sl
     Phys. Rev. D} {\bf 54}, 2447 (1996);  M. S. Turner, {\sl
     Phys. Rev. D} {\bf 44}, 3737 (1991); B. Paczynski and
     T. Piran, {\sl Astrophys. J.} {\bf 364}, 341 (1990);
     D. Langlois and T. Piran, {\sl Phys. Rev. D} {\bf 53}, 2908
     (1996); M. Jaroszynski and B. Paczynski, {\sl
     Astrophys. J.} {\bf 448}, 488 (1995);

\bibitem{dipolefour} D. J. Fixsen et al.,
     {\sl Astrophys. J.} {\bf 420}, 445 (1994).

\bibitem{dmr} A. Kogut et al., {\sl
     Astrophys. J. Lett.} {\bf 646}, L5 (1996).

\bibitem{italians} P. de Bernardis et al., {\sl Astrophys. J.} {\bf
     353}, 145 (1990); S. Bottani, P. de Bernardis, and
    F. Melchiorri, {\sl Astrophys. J. Lett.} {\bf 384}, L1
    (1992).

\bibitem{audit} E. Audit and J. F. L. Simmons,
     {\sl Mon. Not. Roy. Astron. Soc.} {\bf 305}, L27 (1999);
     S. Y. Sazonov and R. A. Sunyaev, {\sl
     Mon. Not. Roy. Astron. Soc.} {\bf 310}, 765 (1999);
     A. D. Challinor, M. T. Ford, and A. N. Lasenby, {\sl
     Mon. Not. Roy. Astron. Soc.} {\bf 312}, 159 (2000);
     D. Baumann, A. Cooray, and M. Kamionkowski,
     astro-ph/0208511.

\bibitem{challinortwo} A. Challinor and F. van Leeuwen,
     astro-ph/0112457.

\bibitem{fixsen98} D. J. Fixsen et al., {\sl Astrophys. J.} {\bf
     508}, 123 (1998). 

\bibitem{tegmark00} M. Tegmark, D. J. Eisenstein, W. Hu, and
     A. de Oliveira-Costa, {\sl Astrophys. J.} {\bf 530}, 133 (2000). 

\bibitem{springel} V. Springel, M. White, and L. Hernquist,
     {\sl Astrophys. J.} {\bf 549}, 681 (2001).

\bibitem{kogut96} A. Kogut et al., {\sl Astrophys. J.} {\bf 460}, 1 (1996).

\bibitem{pw} P. J. E. Peebles and D. T. Wilkinson, {\sl
     Phys. Rev.} {\bf 174}, 2168 (1968). 

\end{thebibliography}
\end{document}